# Photonic-chip assisted correlative light and electron microscopy

Jean-Claude Tinguely#, Anna Maria Steyer#, Cristina Ionica Øie, Øystein Ivar Helle, Firehun Tsige Dullo, Randi Olsen, Peter McCourt, Yannick Schwab, Balpreet Singh Ahluwalia*

# Authors with equal contribution

*Corresponding author:* [Balpreet.singh.ahluwalia@uit.no](mailto:Balpreet.singh.ahluwalia@uit.no)

**Abstract**: *Correlative light-electron microscopy (CLEM) unifies the versatility of light microscopy (LM) with the high resolution of electron microscopy (EM), allowing one to zoom into the complex organization of cells. Most CLEM techniques use ultrathin sections, and thus lack the 3D-EM structural information, and focusing on a very restricted field of view. Here, we introduce photonic chip assisted CLEM, enabling multi-modal total internal reflection fluorescence (TIRF) microscopy over large field of view and high precision localization of the target area of interest within EM. The chip-based direct stochastic optical reconstruction microscopy (dSTORM), and 3D high precision correlation of biological processes by focused ion beam-scanning electron microscopy (FIB-SEM) is further demonstrated. The core layer of the photonic chips are used as a substrate to hold, to illuminate and the cladding layer is used to enable high-precision landmarking of the sample through specially designed grid-like numbering systems. The landmarks are fabricated on the cladding of the photonic chips as extruding pillars from the waveguide surface, thus remaining visible for FIB-SEM after resin embedding during sample processing. Using this approach we demonstrate its applicability for tracking the area of interest, imaging the 3D structural organization of nano-sized morphological features on liver sinusoidal endothelial cells such as fenestrations, and correlating specific endo-lysosomal compartments with its cargo protein upon endocytosis. We envisage that photonic chip equipped with landmarks can be used in the future to automatize the work-flow for both LM and EM for high-throughput CLEM, providing the resolution needed for insights into the complex intracellular communication and the relation between morphology and function in health and disease.*

The photonic-chip consisting of optical waveguides was recently introduced as a platform to perform large field of view (FOV) multi-modal optical nanoscopy[1, 2]. The sample is hosted directly on top of a waveguide chip and is illuminated by the evanescent field present on top of the waveguide surface enabling on-chip total internal reflection fluorescence (TIRF) microscopy. The light into the waveguide is coupled from the side facet of the chip and the fluorescence signal from the sample is collected by any standard optical microscope. This configuration decouples the dependence between the excitation and the collection optics, enabling adjustment-free wavelength multiplexing and free choice of imaging objective without influencing the illumination light path. High intensities (1-10 kW/cm$^2$) in the evanescent field can be achieved by fabricating waveguides made of high-refractive index contrast (HIC) materials (e.g. silicon nitride, $Si_3N_4$) and by using thin waveguide geometry (150-220 nm thickness). The use of HIC materials also enables tight confinement of the light inside the photonic-chip. Moreover, the spatial frequencies of the evanescent field illumination are determined by the refractive index of the material (n=2.05 for $Si_3N_4$) and are thus higher than can be generated even by the oil immersion objective lens (n=1.49). Thus, chip-based microscopy/nanoscopy provides a unique opportunity of imaging large FOV supported by low magnification objective lens while keeping the resolution supported by the high spatial frequencies supported by the HIC waveguide material. These properties were harnessed for multi-modal nanoscopy, including chip-based implementations of *d*STORM (direct stochastic optical reconstruction microscopy)[1,3], TIRF-structured illumination microscopy (TIRF-SIM)[4], and light fluctuation based optical nanoscopy, such as ESI (entropy based super-resolution imaging)[1] (See Supplementary Information "Chip-based nanoscopy"). Notably, chip-based *d*STORM demonstrated an optical resolution of 72 nm over an extraordinarily large FOV, 0.5 mm x 0.5 mm, and chip-based TIRF-SIM showed resolution enhancement of 2.4X which surpassed the resolution supported by conventional TIRF-SIM[5].

Here, we extend the utility of the photonic-chip illumination towards 3D correlative light electron microscopy (CLEM). This is demonstrated by combining chip-based TIRF and *d*STORM with FIB-SEM to demonstrate wide-field imaging, super-resolution nanoscopy and 3D-EM for cell biology, using a single photonic-chip. Correlative light and electron microscopy (CLEM) are widely used techniques, from light microscopy on living samples[6] to on-section CLEM[7]. It has been performed on chemically fixed[8], as well as frozen[9] samples and scales from virus[10] up to cells in mouse brain[11]. Light microscopy (LM) benefits from the availability of a large variety of contrast mechanisms to collect functional, dynamic and specific information using targeted fluorescence labelling methods. Moreover, LM has a larger FOV as compared to EM enabling screening of large sample sets ensuring capturing of localized events such as rare events or addressing heterogeneous subpopulations. On the other hand, EM brings complementary high-resolution down to nanoscale. For the best results in CLEM experiments, data from different modalities should be registered with the highest possible precision. This is normally achieved using reference points visible in both modalities, i.e. fiducial markers[12], which is cumbersome due to challenges related to the location position of the point of reference from one microscopy platform to another, and due to the fluorescence loss upon the use of chemical fixation and heavy metals during EM sample preparation.

To overcome the challenges related to the point of reference location, we designed a special grid-like numbering (landmarks) on the surface of the entire photonic-chip (Figure 1A and Supplementary Information "Chip fabrication") in the form of silicon dioxide pillars (cladding layer). While the thickness of the waveguide layer is only 150 nm, the landmark height is 1.5 µm. The aim here is to make the landmark visible during SEM imaging, after the sample is further stabilized by a thin layer of resin, needed for processing in the FIB-SEM (Figure 1C). This allows precise and fast localization of the cell of interest in the SEM (Supplementary Information "Landmark visibility").

The detailed chip fabrication process can be found in the Supplementary Information "Chip fabrication". The thickness of stripe waveguides made of Si$_3$N$_4$ material was 150 nm and the width ranging from 10-1000 µm. Between waveguides, a 300 nm tall absorbing reduces stray light from neighbouring structures. A cladding of 1.5 µm SiO$_2$ shields the guiding structures from absorbing impurities close to the input. At the imaging area, the cladding was removed to seed the cell directly on top of the waveguide surface. In the imaging region, a coordinate (numbering) system out of cladding material is designed on top of the absorbing layer (Figure 1A, Supplementary Information "Chip fabrication" and "Landmark visibility"). Each point of these landmarks consists of a square with 5 µm side length followed by a letter for a specific waveguide and a continuous numbering system. Each number is 10 pixels in size with a physical pixel size of 2 µm. The photolithography enables global landmarking across the entire 4-inch wafer with a high fabrication accuracy of 1 µm. The experimental set-up to perform chip-based optical nanoscopy is discussed in the methods section and in the Supplementary Information "Chip-based nanoscopy".

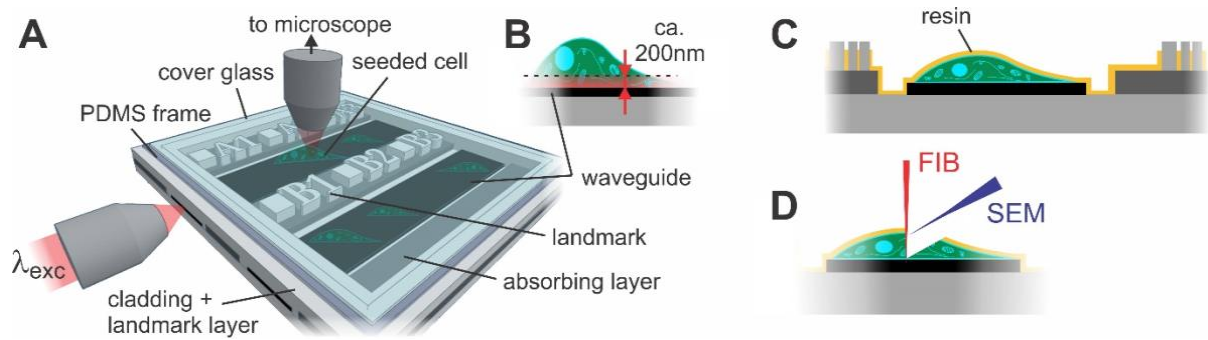

Figure 1: A) Schematic of chip-based TIRF/*d*STORM platform during fluorescence imaging. The samples are placed on top of a waveguide and are illuminated by the evanescent field present on top of the waveguide surface. Different layers of the waveguide chip including the landmark system are shown. B) The evanescent field of the utilized waveguides reaches approximately 200 nm into the sample. C) After light microscopy, a thin layer of resin (in yellow) stabilizes the specimen while maintaining landmark visibility for quick position retrieval inside the focused ion beam - scanning electron microscope (FIB-SEM). D) Volume imaging of cell placed on waveguide chip is performed with FIB-SEM.

To demonstrate the photonic-chip based CLEM methodology and its applicability in cell biology, we used primary rat liver sinusoidal endothelial cells (LSECs). These cells are a very specialized type of endothelial cells, with unique morphology and function, and represent a perfect tool for nanoscopy[13]. Their extensive cytoplasm contains many trans-membrane pores (fenestrations) of approximately 50–200 nm in diameter, typically grouped together in so called sieve plates[14, 15]. In a normal liver, the fenestrations function as a sieve, retaining the blood cells in the sinusoidal lumen, and allowing molecules smaller than the size of the fenestrae, such as metabolites, plasma proteins, pharmaceutical drugs, lipoproteins and small chylomicron remnants, viruses and exosomes to pass through and access the underlying hepatocytes[15, 16]. Besides their filtration role, LSECs have an extraordinary endocytic capacity, effectively scavenging the blood from a variety of physiological and non-physiological waste macromolecules and nanoparticles[17]. The cells recognize and internalize circulating macromolecules, which are then rapidly trafficked and efficiently degraded in the endo-lysosomal compartment. Here, we used the photonic chip to demonstrate its applicability for visualization and correlation of the LSEC fenestrations and endocytosis using TIRF and *d*STORM, and subsequently, FIB-SEM.

To this end, LSECs, which are adherent cells, were seeded directly onto the photonic chips, and incubated with formaldehyde-treated serum albumin (FSA), a well characterized protein for the endocytosis in LSECs[18], which we tagged with AF647 for LM visualization. Two hours post incubation, the cells were fixed and prepared for TIRF/*d*STORM imaging. Knowing that the next step is FIB-SEM, where the best ultrastructural preservation is needed, cell permeabilizing reagents were not used. We were thus restricted to the use of dyes compatible with fixation and that can penetrate the phospholipid layer of the membrane or enter the cell without permeabilization of the membrane envelope, i.e. the actin stain (Phalloidin 488) and the membrane stain (CellMask Orange™). Further, a resin with appropriate viscosity (90 % Durcupan) as applied in the final infiltration step (see Supplementary Information "Work flowchart"), which was removed as much as possible before polymerisation in order to leave the landmarks on top of the photonic-chips visible[19], was used (Figure 2E, Supplementary Information "Landmark visibility"). Additionally, we needed to optimize the sample preparation in order to preserve the ultrastructural features undamaged. This was done by using a Ted Pella microwave with a temperature control unit (Cold Spot) in the EM processing protocol[10] to allow cooling down the sample in order to avoid overheating the cells while processing. Using these two approaches combined, the cell of interest is approached from the top for electron microscopy while the landmarks are showing through the resin layer (Figure 1D).

As a result of imaging using chip-based TIRF and *d*STORM, followed by FIB-SEM of selected cells of interest, we were able to visualize and correlate in 3D the structural organization of the fenestrations on LSECs, and to correlate the endocytosed cargo with its specific endo-lysosomal compartments (Figure 2 and Figure 3). Figure 2 shows the landmarked waveguide carrying the LSECs, where multi-color chip-based TIRF images were acquired to visualize plasma membrane (in green), endocytic vesicles (in red), and actin filaments (in blue). Brightfield and fluorescence signals were used to scan at low magnification (4x or 20x) for targeting highly endocytic cells showing most intact cytoplasm, see Supplementary Information "Landmark visibility". After locating a cell of interest, brightfield (Figure 2A), chip-based TIRF (Figure 2B-C, Figure 3B, 3E-F) or chip-based *d*STORM images (Figure 3D) were taken at higher resolution simply by changing the collection objective to 60x, 1.2 NA (Figure 3A-C, Supplementary Information "Large FOV and multiplexing"). Upon the completion of the LM imaging, the cover glass and imaging chamber were removed and the sample thoroughly fixed using EM methodology (see Supplementary Information "Work flowchart"). The landmarking enables us to target very precisely individual cells and acquire close to isotropic datasets of the cells with a pixel size of 5 nm x 5 nm x 8 nm (Figure 3, Supplementary movie S1). Figure 3 shows that the structures identified by light microscopy (Figure 3 A, E, H) could be correlated to endosomes/lysosomes and fenestrations in the electron microscopic images. This was visualized after thresholding in Amira (FEI) and looking at the x/z surface (for fenestrations) and at the x/y slicing plane (for the internal organelles). The endocytosed FSA visualized by TIRF (red in Figure 3E) was correlated to its internalization compartments, the endosomes and lysosomes (dark blue and light blue in Figure 3H), while the non-resolved dark-green areas on the cell membranes seen in TIRF (Figure 3E) shown to be the sieve plates containing dozens of fenestrations, as demonstrated and correlated after performing FIB-SEM (Figure 3G). From Figure 3G and 3H one could appreciate that the endocytic processes only occurs in the areas between the fenestrated sieve plates. This is most likely due to the lack of sufficient cytoplasmic volume within the sieve plates to accommodate endocytic machinery such as clathrin coated pits, endosomes and lysosomes. This is further supported by the difference in membrane thickness in the fenestrated area ($56 \pm 19$ nm) as compared to the size of the coated pits/endocytic invaginations ($130 \pm 18$ nm), and endosomes ($409 \pm 128$ nm), or lysosomes ($533 \pm 228$ nm). This may explain the findings of Simon-Santamaria et al.[20] who reported that increased LSEC porosity/fenestration resulted in reduced LSEC endocytosis.

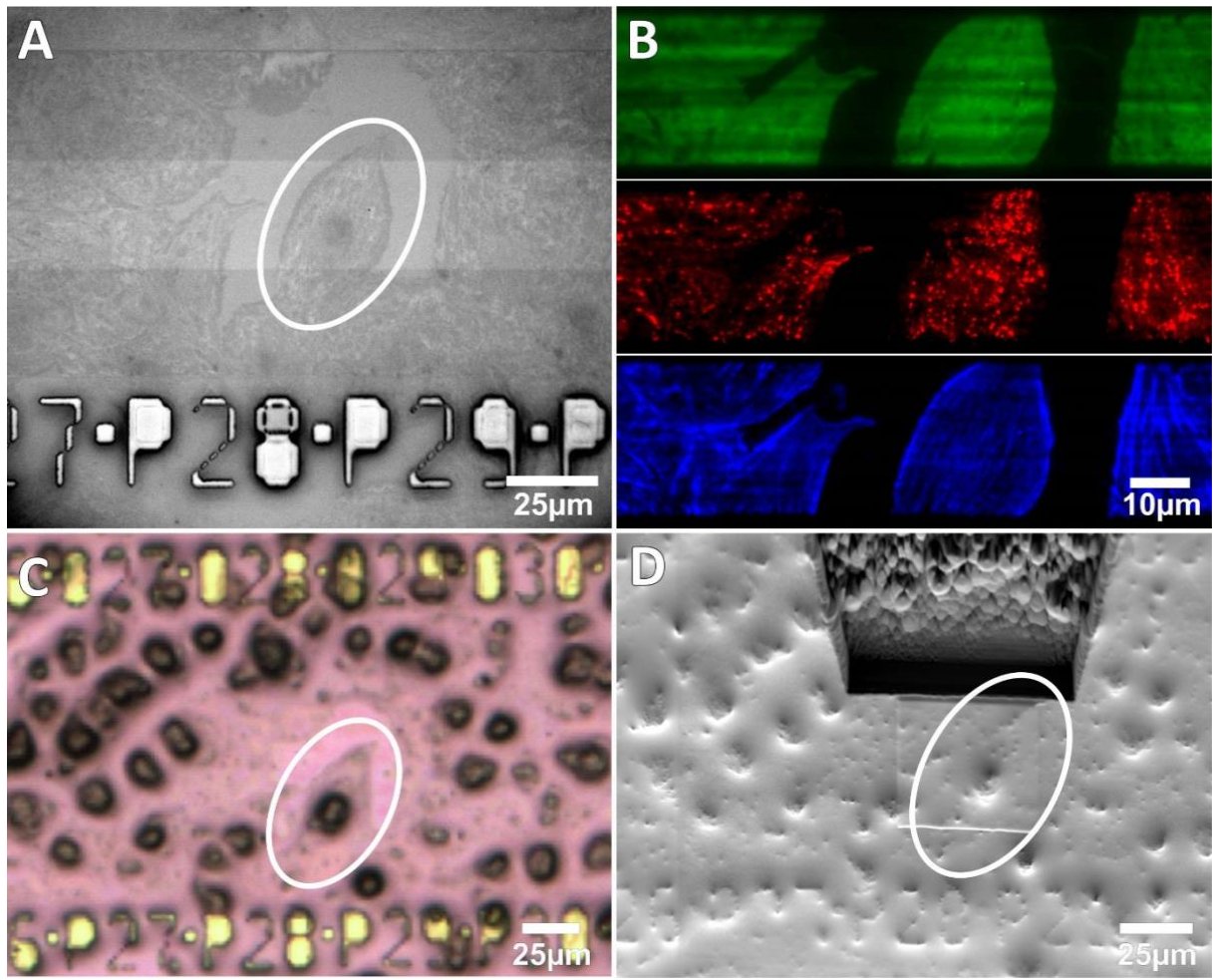

Figure 2: A) Brightfield image of cell of interest with visible landmarks. B) chip-based TIRF image of cell membrane (green), protein undergoing endocytosis (red) and actin (blue). C) Brightfield image after resin coating. D) SEM image of sample processed in the FIB-SEM with visible landmarks.

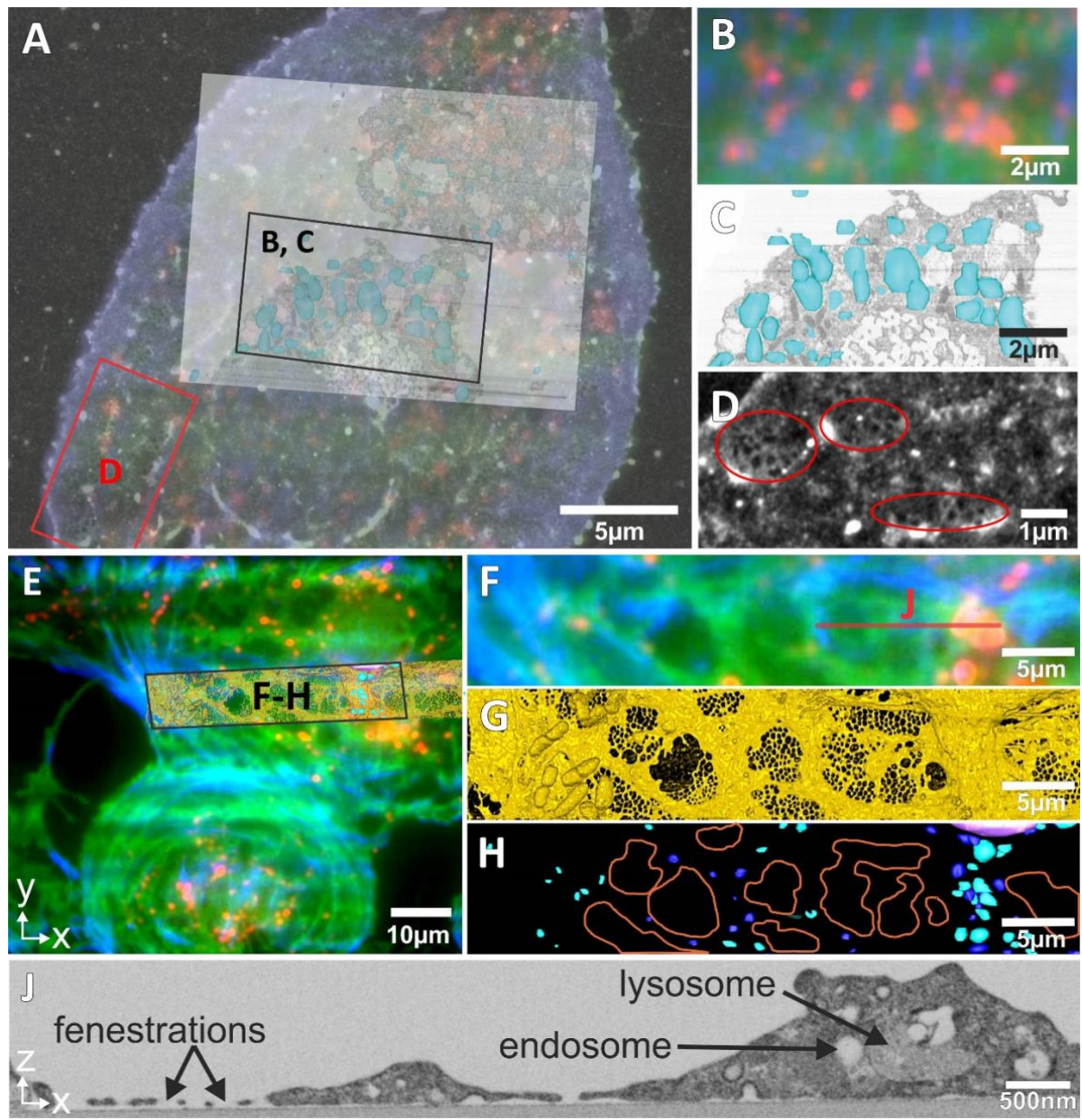

Figure 3: CLEM of LSECs. A) Chip-based TIRF and *d*STORM overlay of an entire LSEC including SEM correlation. B, E, F) Three channel chip-based TIRF, labeling actin (Phalloidin, in blue), cell membrane (CellMask, in green) and FSA after 2 h endocytosis (A-D: AF488-FSA, E-I: AF647-FSA, in red). C) SEM reconstruction of the x/y slicing plane showing lysosomes (light blue). D) chip-based *d*STORM of the cell surface displaying fenestrations. G) SEM reconstruction of the x/z plane of the cell membrane surface showing fenestrations, H) 3D projected overlay of SEM reconstruction showing endosomes (dark blue) and lysosomes (light blue), as well as graphical representation of the sieve plates (orange line). J) FIB-SEM cross-section, x/z plane (Supplementary video 1), corresponding to the red line in F) Comparisons between B) and C) as well as F) and H) indicate the endocytosed FSA being specifically located in the lysosomes.

Imaging on photonics chips opens up a new screening platform for high-resolution light microscopy imaging. Here we have shown that photonic-chip platform can be combined with electron microscopy in a correlative workflow. From the light microscopy side, the photonics chips offer the possibility to acquire not just diffraction limited TIRF imaging, but also *d*STORM on chip, taking again advantage of the huge field of view for screening possibilities. For CLEM, landmark fonts with 2 µm physical pixel size were used. The standard lithographic process can be optimized to allow 1 µm pixel size or less, offering very accurate localisation, while more advanced fabrication methods (such as electron beam writing) can further extend the localization precision. Using microwave-assisted processing and minimal resin embedding allows for a rapid and efficient sample processing, thus fast and easy targeting. 3D-SEM data is then acquired of the same region imaged by nanoscopy to further detail information about three-dimensional internal structures. Conventional CLEM suffers from low throughput, which hinders the collection of sufficient information for drawing relevant conclusions of the biology. However, landmarked photonic-chip technology opens the prospects for automatic targeting of the cells of interest under EM using machine learning of the landmarks. Moreover, the waveguide chip can be pigtailed using an optical fibre (See Supplementary Information "Waveguide setup") to deliver laser illumination directly inside the electron microscope, opening avenues for photonic chip-based integrated CLEM. Photonic chips offer great miniaturization of versatile design sets used to shape and control optical beams that can be barely engineered with conventional far-field optics. Thus, chip-illumination opens possibilities of performing chip-based *d*STORM[4,5], ESI[4], SOFI and TIRF-SIM[6] directly inside the EM.

**ACKNOWLEDGMENTS**
The authors thank Deanna Wolfson for the help with sample handling. B.S.A acknowledge the funding from the European Research Council (Grant Nos. 336716 and 789817) and the Research Council of Norway (Nano2021, Grant no. 288565).

**AUTHOR CONTRIBUTIONS**
B.S.A. and Y.S. conceived the project idea and supervised the project. All authors designed the research. J.C.T. performed the TIRF-experiments, assisted with non-biological sample preparation and A.M.S performed electron microscopy. Ø.I.H. and J.C.T built the experimental setup. Ø.I.H performed on-chip *d*STORM experiments. J.C.T, A.M.S and C.I.Ø coordinated the CLEM experimental pipeline. C.I.Ø. isolated the cells, and stained and prepared the biological samples. C.I.Ø and P.M designed the biological experiments. R.O. assisted with the sample preparation for EM. J.C.T and A.M.S reconstructed the images, analyzed the data, and created the figures. F.T.D and J.C.T designed and characterized the photonic-chip. J.C.T and A.M.S mainly wrote the paper and all authors contributed to selected sections of the manuscript.

**COMPETING FINANCIAL INTERESTS**

B.A.S. applied for a patent for chip-based optical nanoscopy. The other authors declare no competing financial interest.

**Methods**

**Cell extraction** Rat liver sinusoidal endothelial cells (LSECs) were prepared by collagenase perfusion of the liver, low speed differential centrifugation and Percoll gradient sedimentation[21], followed by the depletion of Kupffer cells (KC) by seeding the nonparenchymal fraction onto plastic culture dishes. The KCs attach more rapidly, thereby enriching the suspension with LSECs.

**Sample preparation** 150 µm thick PDMS (Sylgard 184, Dow Corning) was prepared by spin coating in a Petri dish, and square frames of approximately 1.5-2 cm side length were cut and deposited on the waveguide chips. After coating the area within the PDMS frame with fibronectin, the cells were seeded and incubated for 1 h at 37°C. Non-attached cells were removed by washes with PBS, and the cells incubated for another hour. For endocytosis, cells were incubated with fluorescently-labelled formaldehyde-treated serum albumin (AF647-FSA, 50 µg/ml) for 15 minutes at 37°C. Unbound AF647-FSA was washed off with PBS and the cells incubated at 37°C for 2 h. The cells were washed once and fixed with 2.5 % glutaraldehyde, 4 % formaldehyde and 0.05 % malachite green in 0.1 M cacodylate buffer for 15 min at RT. The plasma membrane was stained by incubating the cells for 10 min at RT with CellMask$_{TM}$ Orange 561 (1.25 ng/ml in PBS). The actin filaments were stained by incubating cells for 45 min at 37°C with Alexa Fluor 488 Phalloidin (1:40 dilution in PBS). For the imaging buffer, 22.5 µl of a $H_2O$-based oxygen scavenger system solution (based on glucose oxidase and catalase, Sigma-Aldrich) was mixed with 30 µl PBS[22]. For *d*STORM measurements, 2 mM cyclooctatetraene and 95 mM mercaptoethylamine (Sigma-Aldrich) were added to the imaging media[1]. After adding the imaging media to the cell chamber, the chamber was sealed by gently pressing down a coverslip against the PDMS frame.

**Waveguide imaging.** The waveguides with the fixed cells were imaged on a custom-made system. Three lasers with wavelengths at 488 nm (Oxius), 561 nm (Cobolt) and 660 nm (Cobolt) had their beam width increased with two lens telescope systems to fill the back aperture of the in-coupling objective (Olympus 50x, 0.5 NA). Optical elements as neutral density filters to reduce the minimum laser power were used when necessary. The in-coupling objective and the sample holder were mounted on two separate multi axis stages (Thorlabs), the coupling objective stage disposing of piezo motors and the sample stage of a vacuum mount (more details and schematic diagram under Supplementary Information "Waveguide setup"). For the imaging of specimens, a square shaped polydimethylsiloxane (PDMS) frame (Sylgard 184, Down Corning) of 150 µm thickness was positioned around the imaging area. After placing the sample at the vacuum mount, laser light was focused onto the waveguide edge with the coupling efficiency optimized through the piezo motors. Brightfield and fluorescence images from the waveguide surface were collected by 4x, 20x and 60x (1.2 NA, water immersion) objectives (Olympus) at a modular microscope system (Olympus BXFM). The microscope equipped with a filter wheel (Thorlabs, through home-made adaptor) with a notch and a long pass filter (AHF) for each laser wavelength, where the magnification-free tube directed the light to a sCMOS camera (Orca-Flash 4.0, Hamamatsu). As the waveguides are multimode at the excitation wavelengths, the illumination pattern by the coupling objective at focal distance is highly inhomogeneous. This interference between the modes can be heavily reduced towards a more homogeneous distribution by oscillating the coupling objective along the input facet of the waveguide (see Supplementary Information "Homogeneous illumination with multimode waveguides"). After imaging, the coverslip and PDMS frame were removed and the chips submerged again in fixative containing 2.5 % glutaraldehyde, 4 % formaldehyde, 0.05 % malachite green in 0.1 M cacodylate buffer.

Waveguide processing for electron microscopy. All processing was done in a Ted Pella microwave with a temperature control unit (Ted Pella Inc., USA). Because the photonic-chips were overheating upon microwave processing, the samples were placed directly on the control unit, set to 4°C, and the vacuum chamber was inverted on top. The cells were fixed for 14 minutes (2 min vacuum on-off-on-off-on-off-on, 100 W) and washed two times with 0.1 M cacodylate buffer. Post-fixation was done with 1 % Osmium tetroxide, 1 % $K_3Fe(CN)_6$ in 0.1 M cacodylate. The cells were post-stained with 1 % tannic acid and 1 % uranyl acetate. Samples were then dehydrated in increasing ethanol series (30 %, 50 %, 75 %, 90 % and 2x 100 %) and embedded in increasing amounts of Durcupan (30 %, 50 %, 75 %, 90 %). To be able to remove as much resin as possible, the resin exchange steps went up to 90 % Durcupan in EtOH and not 100 % Durcupan. The chip was centrifuged for 30 minutes at 37°C to further remove excess resin and polymerized in the oven for 96 h at 60°C. The chips were then cut to a final size of 1 cm² to fit the SEM stubs.

**FIB-SEM on cell-monolayer.** The photonics chip with minimally embedded cells and landmarks on the surface was mounted on a SEM stub (6 mm length, Agar Scientific) using a conductive carbon sticker (12 mm, Plano GmbH, Germany) To reduce the amount of charging, the samples were surrounded by silver paint and gold coated for 180 sec at 30 mA in a sputter coater (Quorum, Q150RS). The samples were introduced into the Crossbeam 540 (Carl Zeiss Microscopy, Germany). Light microscopy images of the landmarks were used to target the correct cell inside the FIB-SEM. The FIB was used at 15 nA to mill a trench and expose a cross-section through the cell. A current of 3 nA was used for polishing the cross-section before imaging. For imaging, the FIB milling was operated with 1.5 nA, the SEM imaging and the FIB milling was operating simultaneously[23]. The SEM images were acquired at 1.5 kV with the Energy selective Backscattered (EsB) detector with a grid voltage of 1100 V, analytical mode at a 700 pA current, setting the dwell time and line average to add up to 1.5 min per image. The final dataset was acquired with 5 nm x 5 nm pixel size and a slice thickness of 8 nm.

**Image post-processing.** The following image post-processing steps were performed in Fiji[24]. The image stacks were aligned using TrackEM2[25], cropped and inverted. Following a smoothing step, the 3D segmentation of the fenestration was performed using thresholding in Amira (Thermo Fisher). The individual organelles were segmented in IMOD[26].

**Supplementary Note "Chip-based nanoscopy"**

Total internal reflection fluorescence (TIRF) microscopy makes use of evanescent field-based illumination to limit the excitation and detection of fluorophores to a thin region of the specimen. Typically generated using a specialized TIRF objective lens at the surface of a glass coverslip, the evanescent field will reach to about approximately 100-200 nm into the sample from the surface. The evanescent field excitation significantly improves the axial spatial resolution by only illuminating a thin section. Moreover, by removing fluorescence signal from outside the focal plane, the signal-to-noise ratio is significantly improved. Compared to epi-illumination, the evanescent illumination significantly decreases phototoxicity as well as bleaching of fluorophores[27].

Recently, optical waveguide chips have been introduced as an alternative for large field-of-view (FOV), multimodal TIRF microscopy[2] and optical nanoscopy platform[1, 3, 4]. The light coupled into an optical waveguide is totally internal reflected at the interface boundary of the core-cladding. A part of the guided light is made available on top of the waveguide surface as non-propagating evanescent field. This evanescent field is harnessed to illuminate the sample placed on top of the waveguide surface. By using high-refractive index contrast waveguide material and thin waveguide geometry, high intensity in the evanescent field on top of waveguide surface can be achieved. Waveguides made of silicon nitride and tantalum pentoxide have been demonstrated to achieve a surface intensity of about 1-10KW/$cm^2$, allowing for single molecule localization (SLM) techniques such as *d*STORM (direct stochastic optical reconstruction microscopy) to perform super-resolution imaging through evanescent fields which reach approximately 200 nm into specimens. By using *d*STORM on a chip, a resolution of 75 nm over a field of view of 500 x 500 $\mu m^2$ was reported[3].

**Supplementary Information "Landmark visibility"**

Landmarks were used as a coordinate system in optical (Figure S2A-C) and electron microscopy (Figure S2D-E). Figure S2A-C shows the high contrast of landmarks under brightfield illumination at 4x, 20x, and 60x magnification, respectively. High contrast and thus visibility allow for easy position readout. Figure S2D-E are scanning electron microscopy images of different chips after the resin step, showing resin thickness variations that can occur after centrifugation and polymerization. Despite different degrees of visibility, the landmarks can still be identified in the scanning electron microscope.

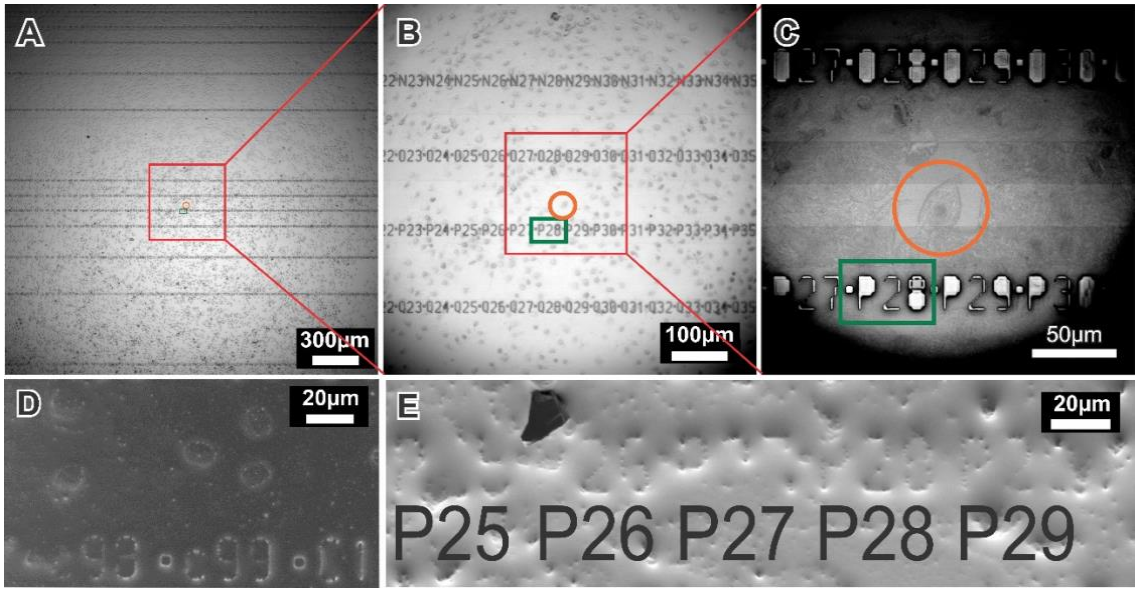

Figure S2A-C: Landmark visibility under optical brightfield microscope. A) 4x, B) 20x and C) 60x magnification. D-E) Resin embedded chips (2 different samples) under the scanning electron microscope.

**Supplementary Information "Chip fabrication"**

The photonic chips were manufactured through standard photolithography at the Institute of Microelectronics Barcelona (IMB-CNM, Spain). The scheme in Figure S3 depicts only the small section of a single waveguide for better visualization. 1 mm thick, 4" silicon wafers were used as substrates (Figure S3A). First, a $SiO_2$ layer with a thickness of approximately 2 µm was thermally grown on the surface (Figure S3B). 150 nm of $Si_3N_4$ were deposited using low-pressure chemical vapour deposition (LPCVD) at 800°C (Figure S3C) followed by thermal annealing at 600°C. Photolithography was employed to pattern the waveguide geometry, with reactive ion etching (RIE) defining the etch depth (Figure S3D). The remaining photoresist was removed, and 200 nm $SiO_2$ followed by 100 nm polycrystalline silicon deposited by plasma-enhanced chemical vapour deposition (PECVD) at 300°C. These 300 nm of material were patterned by RIE (first 200 nm) and hydrofluoric acid (last 100 nm) as channels beside the waveguides (10 µm gap between waveguide and absorbing layer walls) (Figure S3E). Such structures are meant to reduce cross-illumination between neighboured waveguides, of special importance for shallow rib waveguides with low light confinement. For strip etched, multimode waveguides they mostly help towards raising the total height of the landmark structures. Finally, 1.5 µm of $SiO_2$ were deposited for the cladding layer (PECVD). Taller layers were avoided to prevent high stress in the layer and because of the slow deposition process. Photolithography is performed with RIE down to 100 nm of material and hydrofluoric acid for the remaining $SiO_2$ thickness. An area of 13 x 14 mm was so left exposed at the center of the chip with landmark structures placed on the absorbing layer along the direction of light propagation (Figure S3F). The wafers were diced with a precision saw to chips of approximately 2-3 cm side length and polished with abrasive discs of 1 µm and 0.5 µm grit size (Buehler). Optical characterization of 1.5 µm wide waveguides exhibited less than 1dB/cm loss at 660 nm, approximately 2 dB/cm loss at 561 nm and approximately 10 dB/cm loss at 488 nm[2].

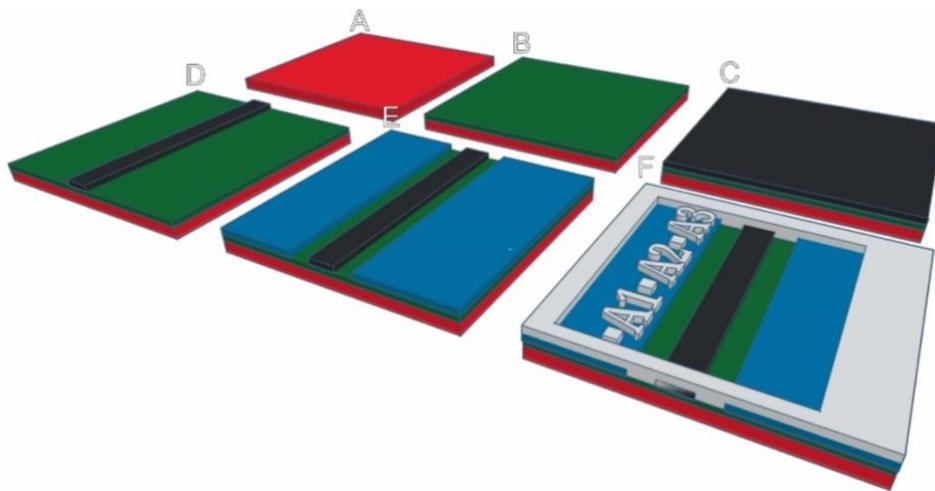

Figure S3: Scheme of fabrication steps of the waveguide chips. A) 1 mm thick silicon wafer. B) Approximately 2 µm thick $SiO_2$ created through thermal oxidation. C) Deposition of 150 nm thick $Si_3N_4$. D) Photolithographic patterning followed by etching of $Si_3N_4$. Waveguide widths for imaging range between 25 to 500 µm. E) Deposition and patterning of absorbing layer consisting of 200 nm thick $SiO_2$ (bottom) and 100 nm thick polycrystalline silicon (on top). F) Deposition and patterning of 1.5 µm thick $SiO_2$ cladding for imaging regions and landmarks.

**Supplementary Note "Endocytosis of FSA by LSECs"**

FSA was tagged with AF647 (red) for visualization. The LSECs seeded on the photonic-chip were incubated with AF647-FSA, 50 µg/ml for 15 minutes at 37°C. Unbound AF647-FSA was washed off with PBS and the cells incubated for an additional 2 h at 37°C. At the end of the incubation time, the cells were fixed. Subsequently, staining for actin using phalloidin (blue) and for cell membrane using CellMask (green) was performed. The chips were then imaged by waveguide-based TIRF. Figure S4: The FSA is seen accumulated in the perinuclear regions in lysosomes.

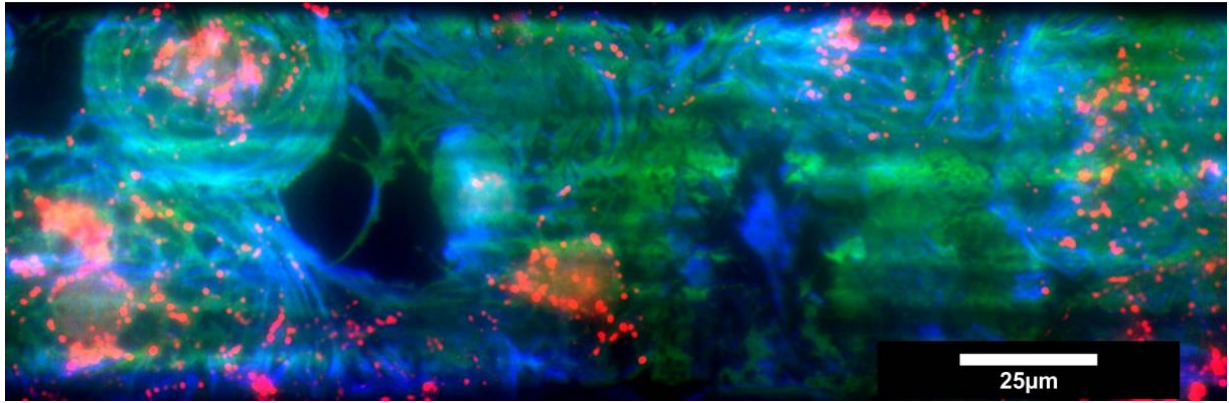

Figure S4: Chip-based TIRF images showing actin (phalloidin, blue), cell membrane (CellMask, green) and the localization of endocytosed FSA (AF647, red) after being endocytosed by the cells.

**Supplementary Note "Work flowchart"**

The flowchart below provides an overview of the processing steps from chip preparation, cell seeding and staining, light microscopy imaging, sample preparation for electron microscopy and FIB-SEM imaging.

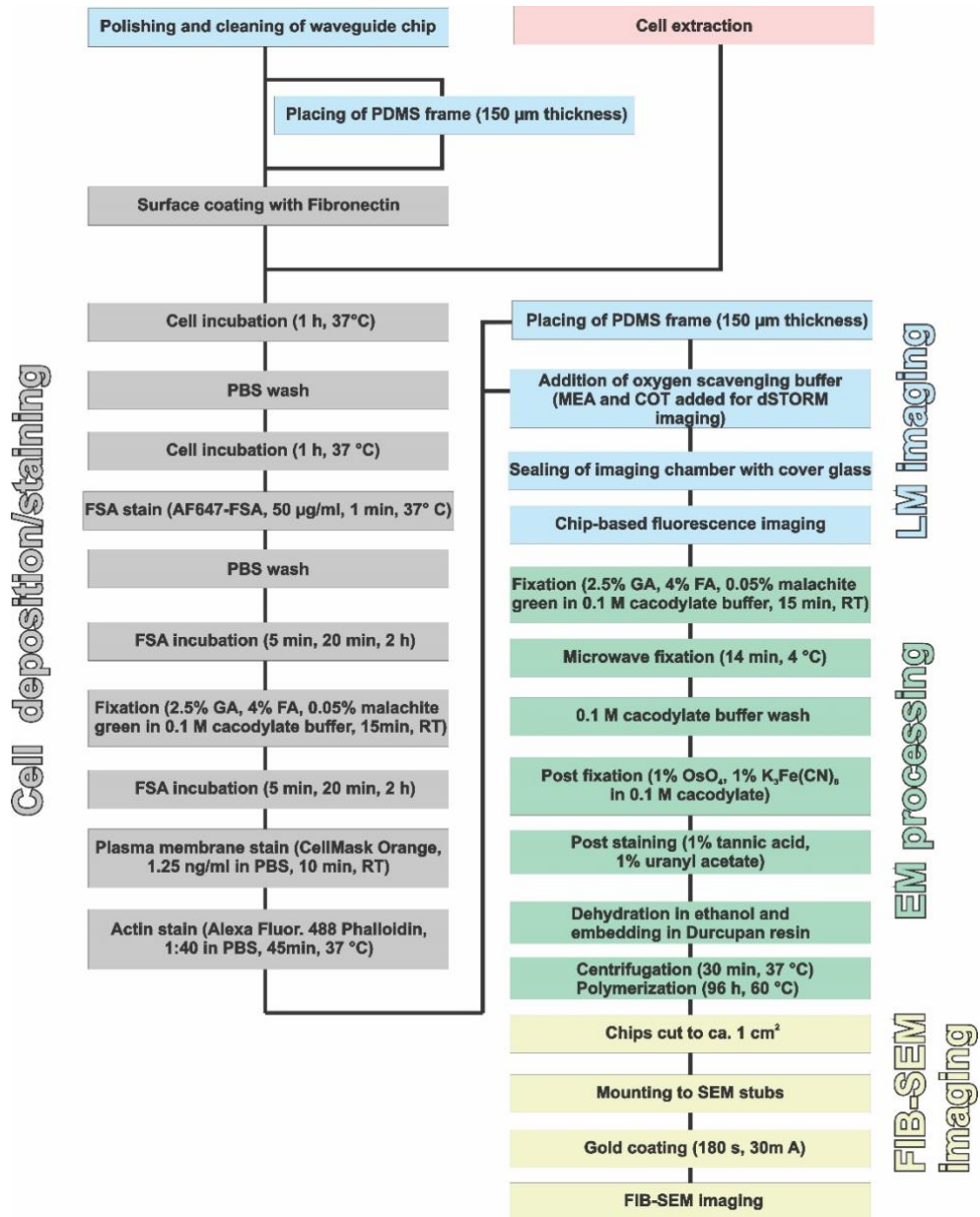

Figure S5: Work flowchart.

**Supplementary Information "Large FOV and multiplexing"**

Making use of a waveguide chip as a substrate and illumination source decouples the typical dependence between excitation and collection optics. It offers generation of evanescent field without the necessity of a specialized high magnification/numerical aperture TIRF lens, enabling free choice of imaging objective. This property is of advantage for large field of view TIRF-imaging applications. Here, an isolated LSEC cell was made out from the membrane stain channel under 4x magnification, after which a 60x image was taken (Fig. S6).

For conventional, objective-based TIRF setups, multicolor imaging is cumbersome as it requires repositioning of the beam to fulfill the total internal reflection condition when changing between excitation wavelengths. The waveguide setup allows for easy multiplexing, as an achromatic coupling lens will provide the same beam spot for coupling between different laser lines without the necessity of mechanical adjustments. As propagation losses in the waveguide are higher for shorter wavelengths, compensation in the illumination intensity might be necessary. It was however observed that the intensity of the fluorophore stain had at least a similar influence in the signal to noise ratio.

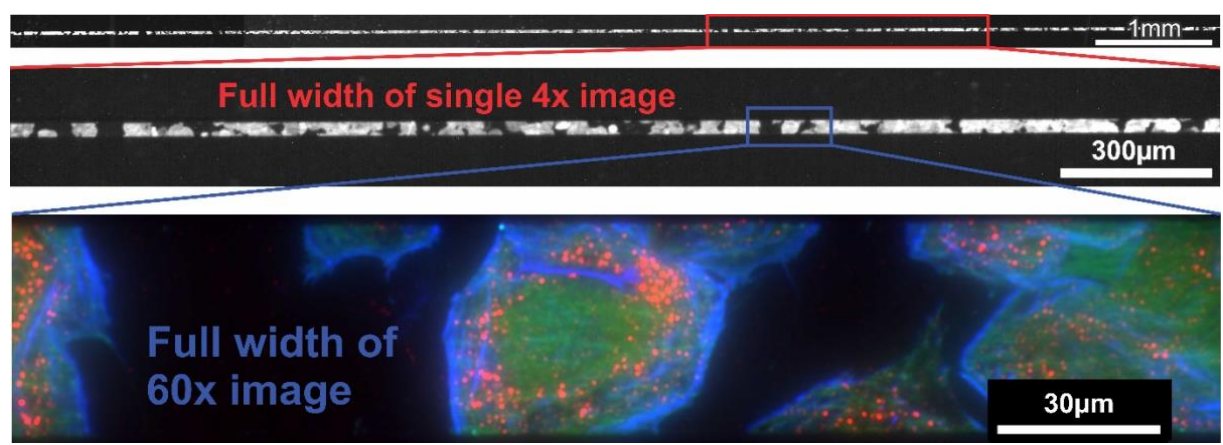

Figure S6: Decoupling of excitation and collection in waveguide-based TIRF imaging allows for simple switching between low magnification overview and high-resolution images. Top image: ca. 1 cm stitched from 4x 0.1 NA images, cell membrane staining. Center image: Full width of single low magnification image. Lower image: Full width of high-resolution image (60x, 1.2 N.A.), three channel waveguide-based TIRF.

**Supplementary Information "Waveguide setup"**

A scheme of the setup is shown in Figure S7 A Laser beams at three conventional bioimaging wavelengths (561 nm 500 mW and 660 nm 500 mW- Cobolt, 488 nm 200 mW - Oxius) are expanded to slightly overfill the back aperture of the coupling objective (Olympus, 50x / 0.5 NA). The laser beams are aligned using dichroic mirrors (Edmund Optics) towards the coupling objective mounted on a 3 axis flexure stage featuring piezo motors (Thorlabs Nanomax). The waveguide chip is mounted on a separate 3 axis flexure stage (Thorlabs Nanomax) with a vacuum chuck (Thorlabs) to hold the chip. A modular microscope setup (Olympus BXFM) mounted on motorized translation stages for two axis translation (Thorlabs) provides brightfield illumination and acquires images at different magnifications (4x, 20x / 0.45 NA, 60x / 1.2 NA water immersion, all Olympus). High magnification is used with a z-piezo system (PIFOC, PI) for fine adjustment. A homemade adapter holds a cage filter wheel (Thorlabs) between the microscope body and the tube (1x). For each excitation wavelength, a long pass and a band pass filters were used (488 LP and 520 ± 36 nm, 561 LP and 591 ± 43 nm, 664 LP and 692 ± 40 nm, AHF Analysentechnik), and the images recorded with an sCMOS camera (Hamamatsu Orca Flash).

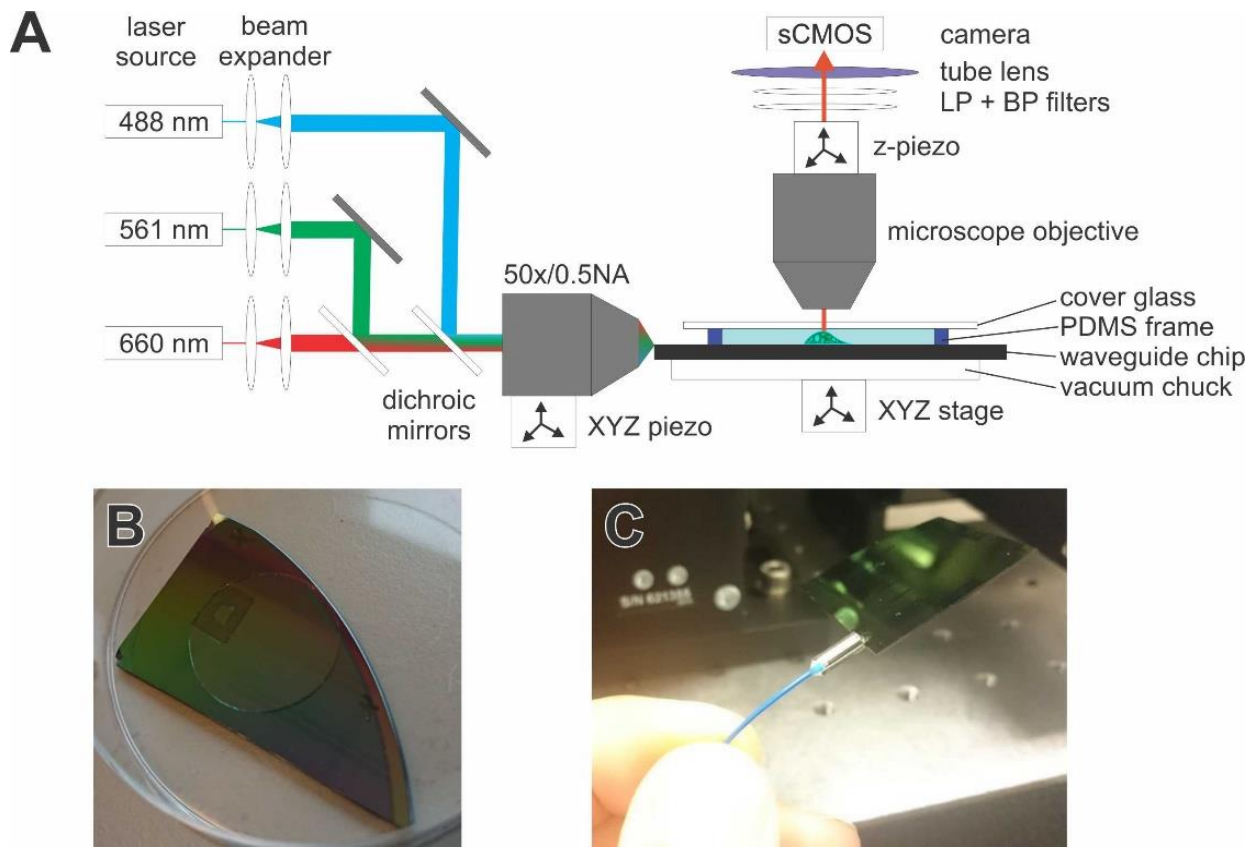

Figure S7: Waveguide setup A) Experimental-set-up for waveguide-based optical micro/nanoscopy. B) Waveguide chip with mounted PDMS frame and coverslip sealing the imaging region. C) Single-mode fiber pigtailed to waveguide facet as an alternative objective-based coupling method, which could be useful for integrated CLEM.

**Supplementary Note "Homogeneous illumination with multimoded waveguides"**

Recently, for $Si_3N_4$ waveguide platform operating at visible wavelengths, a single condition carrying only the fundamental mode was demonstrated using a rib waveguide geometry[3] for 150 nm thick and 1.5 µm wide $Si_3N_4$ waveguides. Adiabatic tapering can increase the width of the structures for imaging, but required taper length for widths of hundreds of micrometers would be of several millimetres[2]. In this work, we employed stripe geometry creating multimoded waveguides. The presence of multiple modes creates interference patterns and so uneven illumination on the surface. This can be reduced by scanning the coupling beam over the waveguide width (Figure S8) and averaging the mode beating pattern generating a uniform image[4]. This allows for the use of waveguide structures of several hundreds of micrometers in width.

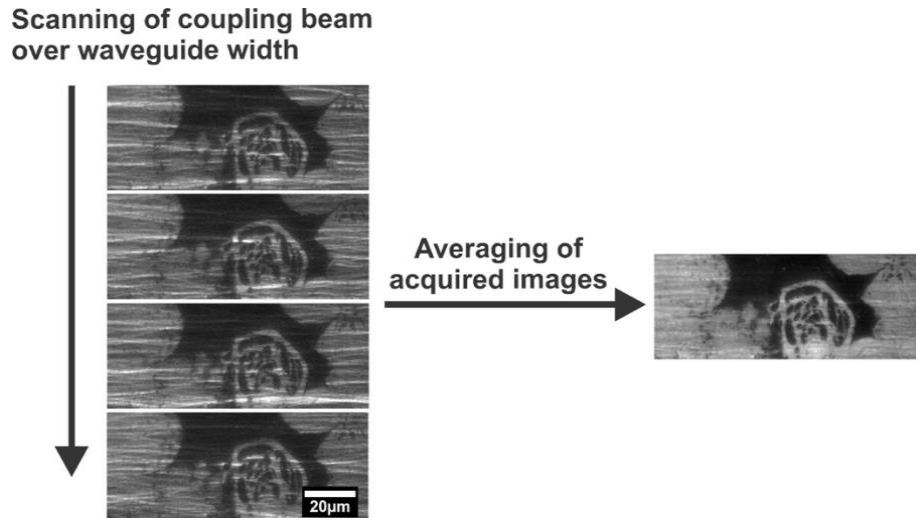

Figure S8: Left: Multimode waveguides provide uneven illumination through interference patterns between modes. These patterns can be varied when moving the coupling beam along the input facet. Right: Averaging images of these mode variations results provide an almost homogeneous illumination pattern.

**Supplementary Video 1. High-Resolution FIB-SEM 3D-reconstruction of fenestrations and endo-lysosomal compartment in one LSEC (Related to Figure 3)**